\documentclass{rspublic}

\makeatletter
\newcounter {paragraph}[subsubsection]
\renewcommand\theparagraph    {\thesubsubsection.\@arabic\c@paragraph}
\newcommand\paragraph{\@startsection{paragraph}{4}{\z@}%
                                    {3.25ex \@plus1ex \@minus.2ex}%
                                    {-1em}%
                                    {\normalfont\normalsize\bfseries}}

\let\bibdata=\@gobble
\let\bibstyle=\@gobble
\def\bibliography#1{%
  \if@filesw
    \immediate\write\@auxout{\string\bibdata{#1}}%
  \fi
  \@input@{\jobname.bbl}}
\def\bibliographystyle#1{%
  \ifx\@begindocumenthook\@undefined\else
    \expandafter\AtBeginDocument
  \fi
    {\if@filesw
       \immediate\write\@auxout{\string\bibstyle{#1}}%
     \fi}}
\def\nocite#1{\@bsphack
  \@for\@citeb:=#1\do{%
    \edef\@citeb{\expandafter\@firstofone\@citeb}%
    \if@filesw\immediate\write\@auxout{\string\citation{\@citeb}}\fi
    \@ifundefined{b@\@citeb}{\G@refundefinedtrue
        \@latex@warning{Citation `\@citeb' undefined}}{}}%
  \@esphack}
\expandafter\let\csname b@*\endcsname\@empty
\def\@cite#1#2{[{#1\if@tempswa , #2\fi}]}
\makeatother

\usepackage{color}
\usepackage{epic}
\usepackage{graphics}
\usepackage[agsm]{harvard}
\usepackage{psfrag}

\allowdisplaybreaks

\renewcommand{\d}{\mathrm{d}}		
\newcommand{\df}[2]{{\partial #1}/{\partial #2}} 	
\newcommand{\ddf}[2]{{\partial^2 #1}/{\partial #2^2}}	
\newcommand{\Df}[2]{{\d #1}/{\d #2}}	
\newcommand{\Eq}[1]{(\ref{#1})} 	
\newcommand{\fig}[1]{fig.~\ref{#1}} 	
\newcommand{\Fig}[1]{Fig.~\ref{#1}}	
\newcommand{\Heav}{\theta}		
\newcommand{\myfigure}[3]{		
\begin{figure}[t]
\resizebox{\textwidth}{!}{\includegraphics{#1}}
\caption[]{\small #2}
\label{#3}
\end{figure}
}

\newcommand{\cm}{\ensuremath{\mathrm{cm}}} 
\newcommand{\mS}{\ensuremath{\mathrm{mS}}} 
\newcommand{\mV}{\ensuremath{\mathrm{mV}}} 
\newcommand{\mm}{\ensuremath{\mathrm{mm}}}
\newcommand{\ms}{\ensuremath{\mathrm{ms}}} 
\newcommand{\uA}{\ensuremath{\mu\mathrm{A}}} 
\newcommand{\uF}{\ensuremath{\mu\mathrm{F}}} 

\renewcommand{\i}{i}			
\newcommand{\n}{N}			
\newcommand{\y}{y}			

\newcommand{\mx}[1]{\mathbf{#1} }	
\newcommand{\F}{\mx{f}}			
\newcommand{\G}{\mx{g}}			
\newcommand{\X}{\mx{X}}			
\newcommand{\Y}{\mx{Y}}			

\newcommand{\emb}[1]{\hat{{#1}}}	
\newcommand{\Qstat}[1]{\overline{#1}}	
\newcommand{\CM}{C_M}			
\newcommand{\EK}{E_K}			
\newcommand{\ENa}{E_{Na}}
\newcommand{\El}{E_l}
\newcommand{\INa}{I_{\mathrm{Na}}}
\newcommand{\dbar}{\Qstat{d}}
\newcommand{\gKi}{g_{K_1}}
\newcommand{\gK}{g_{K}}
\newcommand{\gNai}{g_{Na_1}}
\newcommand{\gNa}{g_{Na}}
\newcommand{\gl}{g_{l}}
\newcommand{\mbar}{\Qstat{m}}
\newcommand{\nbar}{\Qstat{n}}
\newcommand{\oabar}{\Qstat{\oa}}
\newcommand{\oa}{o_a}
\newcommand{\oi}{o_i}
\newcommand{\uabar}{\Qstat{\ua}}
\newcommand{\ua}{u_a}
\newcommand{\wbar}{\Qstat{w}}
\newcommand{\ybar}{\Qstat{\y}}
\renewcommand{\hbar}{\Qstat{h}}

\newcommand{\Ealp}{E_-}			
\newcommand{\Eomg}{E_+}			
\newcommand{\Eh}{E_h}			
\newcommand{\Em}{E_m}			

\newcommand{\jmin}{j_{\min}}		
\newcommand{\cmin}{c_{\min}}		

\newcommand{\SI}{{\Sigma_I}}		
\newcommand{\SIsmall}{{\Sigma_{I}^{\prime}}}	

\begin{document}
\title[Asymptotics of excitability]{Asymptotic properties of mathematical models of excitability}
\author[I. V. Biktasheva et al.]{I. V. Biktasheva$^{1}$, R. D. Simitev$^{2}$, R. Suckley$^{2}$, V. N. Biktashev$^{2,*}$}
\affiliation{
  $^1$ Department of Computer Science, 
  University of Liverpool, Liverpool L69 3BX, UK 
  and
  $^2$ Department of Mathematical Sciences, University of
  Liverpool, Liverpool L69 7ZL, UK.
  $^*$ Corresponding author.
}
\label{firstpage}
\maketitle

\begin{abstract}{
  singular perturbations, action potential, front dissipation
}
We analyse small parameters in selected models of biological
excitability, including Hodgkin-Huxley (1952) model of
nerve axon, Noble (1962) model of heart Purkinje fibres, and
Courtemanche et al. (1998) model of human atrial cells. Some of the
small parameters are responsible for differences in the characteristic
timescales of dynamic variables,  
as in the traditional singular perturbation approaches.
Others appear in a way which makes
the standard approaches inapplicable. We apply this analysis to study the
behaviour of fronts of excitation waves in spatially-extended cardiac
models. Suppressing the excitability of the tissue leads to a decrease in
the propagation speed, but only to a certain limit; further
suppression blocks active propagation and leads to a passive
diffusive spread of voltage. Such a dissipation may happen if a front
propagates into a tissue recovering after a previous wave, e.g. re-entry.
A dissipated front does not recover even when the
excitability restores. This has no analogy in
FitzHugh-Nagumo model and its variants, where fronts can stop and then
start again.  In two spatial dimensions, dissipation accounts for
break-ups and
self-termination of re-entrant waves in excitable
media with Courtemanche et al. (1998) kinetics.
\end{abstract}

\section{Introduction} 

The motivation of this study comes from a series of numerical
simulations of spiral waves \cite{Biktasheva-etal-2003} in a model
of human atrial tissue based on the excitation kinetics of
\citeasnoun{Courtemanche-etal-1998} (CRN).  The spiral waves in this
model tend to break up into pieces and even spontaneously
self-terminate (see \fig{selfterm}).

\myfigure{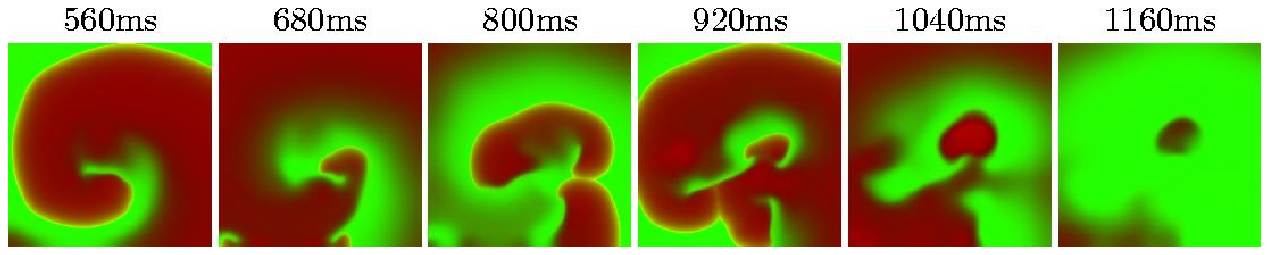}{
Self-termination of a spiral wave in the CRN model. Red colour
component: transmembrane voltage $E$, green
colour component: gating variable $\oi$.
Diffusion coefficient $D=0.03125\mm^2/\ms$, preparation size
$75\times75\,\mm$. 
See also \cite{Biktasheva-etal-2003}.
}{selfterm}

No mathematical model of cardiac tissue is now considered
ultimate or can claim absolute predictive power. 
The spontaneous self-termination may be relevant
to human atrial tissue or may be an artefact
of modelling. 
Understanding
the mechanism of this behaviour in some simple terms
would allow a more direct and certain verification. This is difficult
as the models are very complex and the events depicted in
\fig{selfterm} have many different aspects.  Traditionally,
such understanding has been achieved in terms of simplified models,
starting from axiomatic cellular automata description
\cite{Wiener-Rosenblueth-1946} through to simplified PDE models
\cite{FitzHugh-1961,Nagumo-etal-1962} which allow asymptotic study by
means of singular perturbation techniques \cite{Tyson-Keener-1988}.
This approach can describe some of the features observed in
\fig{selfterm}, e.g. the ``APD restitution slope 1'' theory predicts when the
stationary rotation of a spiral wave is unstable against
alternans
\cite{Nolasco-Dahlen-1968,Karma-etal-1994,Karma-1994}. The relevance
of the ``slope-1'' theory to particular models is debated
\cite{Cherry-Fenton-2004}, but in any case it
only predicts the instability of a spiral wave, not whether
it will lead to complete self-termination of the spiral wave, its
breakup, or just meandering of its tip. 
We need to understand how the propagation of a wave is blocked. 
This has unexpectedly turned out to be rather interesting.
Some features of the propagation block in \fig{selfterm} can
\emph{never} be explained within the standard FitzHugh-Nagumo
approach. As this was the only well developed asymptotic approach to
excitable systems around, we had either to accept that this problem
is too complicated to be understood in simplified terms, or to develop
an alternative type of simplified model and corresponding
asymptotics.

We chose the latter.
This paper summarizes our progress in this direction in the last few
years \cite{%
  Biktashev-2002,%
  Biktashev-2003,%
  Biktasheva-etal-2003,%
  Suckley-Biktashev-2003,%
  Biktashev-Suckley-2004,%
  Suckley-2004,%
  Biktashev-Biktasheva-2005%
}. The results on the asymptotic structure of the CRN model are published
for the first time.

\section{Tikhonov asymptotics}
\label{sec:Tikhonov}

The standard Tikhonov-Pontryagin singular perturbation theory 
\cite{Tikhonov-1952,Pontryagin-1957} summarized in \cite{Arnold-etal-1994}
is usually formulated in terms of ``fast-slow'' systems  in
one of the two equivalent forms
\begin{equation}
\begin{array}{l}
  \Df{\X}{t}=\F(\X,\Y) \\
  \epsilon\Df{\Y}{t}=\G(\X,\Y)
\end{array}
\Leftrightarrow
\begin{array}{l}
  \Df{\X}{T}=\epsilon \F(\X,\Y) \\
  \Df{\Y}{T}=\G(\X,\Y)
\end{array}
\end{equation}
where $\epsilon>0$ is small, $t$ is the ``slow time'', $T$ is the
``fast time'', $t=\epsilon T$, $\X$ is the vector of ``slow
variables'' and $\Y$ is the vector of ``fast variables''. The theory
is applicable if all relevant attractors of the fast subsystem are
asymptotically stable fixed points. So the variables have to be
explicitly classified as fast and slow, and the system should contain
a small parameter which formally tends to zero although the
original system is formulated for a particular value of that
parameter, say 1.

We consider \citeasnoun{Hodgkin-Huxley-1952} (henceforth referred to as HH)
and \citeasnoun{Noble-1962} (henceforth N62) models. Both can
be written in the same form,
\begin{eqnarray}
\Df{E}{t}&=& - \SI(E,h,m,n)/\CM,				\nonumber \\
\Df{\y}{t}&=& {(\ybar(E)-\y)}/{\tau_{\y}(E)},\quad \y=m,h,n, \label{hh+n62}
\end{eqnarray}
where
\(
   \SI (E,h,m,n) = (\gK n^{4}+\gKi(E))(E-\EK) 
		 + (\gNa m^{3}h+\gNai )(E-\ENa) 
		 +  \gl (E-E_l)
\)
is the total transmembrane current ($\uA/\cm^2$), 
$t$ is time (\ms), 
$E$ is the transmembrane voltage (\mV), 
$E_k$, $k=Na,K,l$ are the reversal potentials of sodium, potassium and leakage currents respectively (\mV),
$\Bar{g}_{k}$ are the corresponding maximal specific conductances ($\mS/\cm^2$), 
$n$, $m$, $h$ are dimensionless ``gating'' variables, 
$C_M$ is the specific membrane capacitance ($\uF/\cm^2$), 
$\ybar(E)$ are the gates' instantaneous equilibrium ``quasi-stationary'' values, and 
$\tau_{\y}(E)$ are the characteristic timescale coefficients of the gates dynamics ($\ms$).  
Definition and comparison of parameters and functions used in
\Eq{hh+n62} for HH and N62 models can be found in
\cite{Suckley-Biktashev-2003}.

To classify the dynamic variables by their speeds, we define empiric
characteristic timescale coefficients, $\tau_{\i}$.  For a system of differential equations
$\Df{x_{\i}}{t} = f_{\i}(x_1,\dots,x_{\n})$, $\i=1,\dots,\n$, we define
$\tau_{\i}(x_1,\dots,x_{\n}) \equiv
\left|\left(\df{f_{\i}}{x_{\i}}\right)^{-1}\right|$.  
The $\tau$'s obtained for $m$, $h$ and $n$ in this way
coincide with $\tau_{m,h,n}$ in \Eq{hh+n62}, and this definition
can be extended to other variables, e.g. $E$ in the case of \Eq{hh+n62}.

\paragraph{Hodgkin-Huxley.}

\Fig{hhtikh}(a) shows how $\tau$'s
change during a typical action potential (AP) in the HH model. The speeds of $E$
and $m$ exchange places during the AP, as do the speeds of $h$ and
$n$, but at all times the pair $(E,m)$ remains faster than the
pair $(h,n)$. This suggests introduction into system \Eq{hh+n62}
of a parameter $\epsilon$ which in the limit $\epsilon\rightarrow0$
makes variables $E$ and $m$ much faster than $n$ and $h$:
\begin{eqnarray}
\epsilon \Df{m}{t} &=& {(\mbar(E)-m)}/{\tau_{m}(E)},\nonumber \\
\epsilon \Df{E}{t} &=& -\SI(E,h,m,n)/\CM,    \nonumber \\
         \Df{h}{t} &=& {(\hbar(E)-h)}/{\tau_{h}(E)},\nonumber \\
         \Df{n}{t} &=& {(\nbar(E)-n)}/{\tau_{n}(E)}.
							\label{hh-22}
\end{eqnarray}

\myfigure{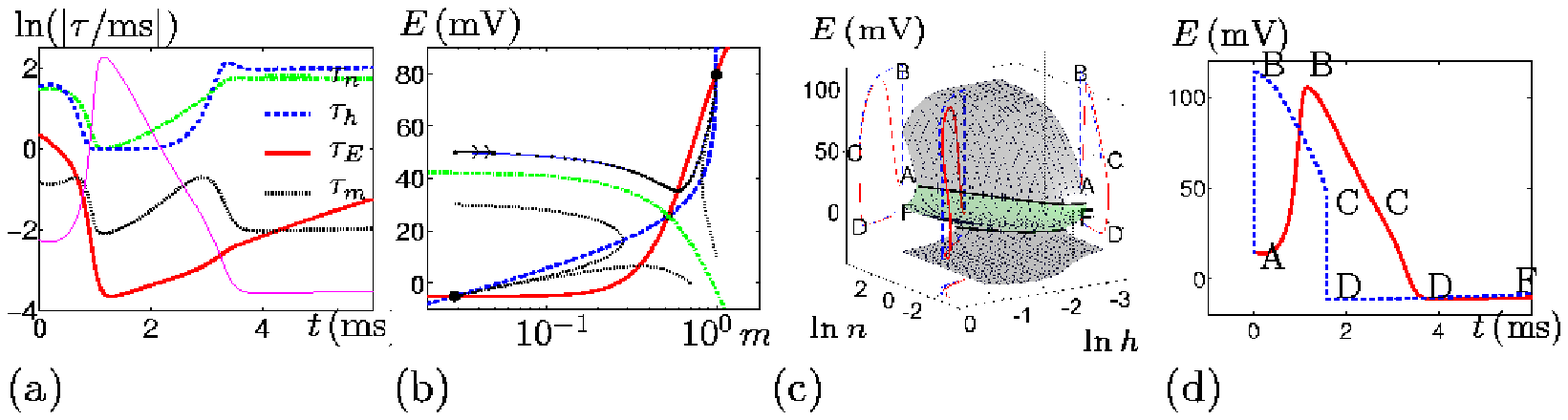}{
Tikhonov asymptotics of HH. 
(a) Characteristic times $\tau_{\i}$ during an AP.
Thin solid magenta line: the AP for a reference.
(b) Phase portrait of the fast subsystem \Eq{hh-fast} 
at $n=0.37$, $h=0.02$.
Solid red line: horizontal isocline.  Dashed blue line: vertical
isocline.  
Filled black circles: stable equilibria.
Dash-dotted green line: stable separatrix of the saddle, 
the boundary of attraction basins. 
Black dotted lines: selected trajectories. 
(c) A three-dimensional view of the slow manifold (the surface). 
Solid line: the fold line.
Dotted line: the selected trajectory and its projections on coordinate walls.
(d) AP in the original model, $\epsilon=1$, solid red line,
and when the fast variables are made faster, $\epsilon=10^{-3}$,
dashed blue line. 
See also \cite{Suckley-Biktashev-2003}.
}{hhtikh}

The properties of this system in the limit $\epsilon\rightarrow0$ are
shown in \fig{hhtikh}(b--c). The fast transient, corresponding to the
AP upstroke can be studied by changing the independent
time variable in \Eq{hh-22} to $T=t/\epsilon$ and then considering the
limit $\epsilon\rightarrow0$ which gives the \emph{fast subsystem} of
two equations for $m$ and $E$,
\begin{eqnarray}
 \Df{m}{T} &=& {(\mbar(E)-m)}/{\tau_{m}(E)},\nonumber \\
 \Df{E}{T} &=& -\SI(E,h,m,n)/\CM, 
						\label{hh-fast}
\end{eqnarray}
in which the slow variables $h$ and $n$ are parameters as their
variations during the onset are negligible. An example of a phase
portrait of system \Eq{hh-fast} at selected values of $h$ and $n$
is shown in \fig{hhtikh}(b). It is \emph{bistable}, i.e it has two
asymptotically stable equilibria, and a particular trajectory 
approaches one or the other depending on the initial conditions. The basins
of attraction\label{basins} of the two equilibria are separated
by the stable
separatrices of a saddle point, which is the threshold between ``all''
and ``none'' responses.  A fine adjustment of initial conditions at
the threshold will cause the system to come to the saddle point. This is a mathematical
representation of the excitation threshold in Tikhonov asymptotics.

For different values of $n$ and $h$, the location of
the equilibria in the fast subsystem vary. All equilibria $(E,m)$
at all values of $n$ and $h$ form a two-dimensional \emph{slow
manifold} in the four-dimensional phase space of \Eq{hh-22} with
coordinates $(E,m,h,n)$. Projection of this two-dimensional manifold
into the three dimensional space with coordinates $(E,h,n)$ is
depicted in \fig{hhtikh}(c). It has a characteristic cubic folded shape,
with two fragments of a positive
slope (as it appears on the figure), separated by an ``overhanging''
fragment of a negative slope. The borders between the fragments are the
\emph{fold lines}, seen as nearly horizontal solid curves on the
picture.  The positive slope fragments consist of
stable equilibria, and the negative slope fragment
consists of unstable equilibria (saddle points) of the fast subsystem.

The points of this manifold are steady-states if considered on the
time scale $T\sim1$ or equivalently $t\sim\epsilon$.  On the time
scale $t\sim1$ these points are no longer steady states, but we
observe a slow (compared to the initial transient) movement along this
manifold, which explains its name. Asymptotically, the evolution on
the scale $t\sim1$ can be described by the limit
$\epsilon\rightarrow0$ in \Eq{hh-22}, which gives a system of two
finite equations and two differential equations,
\begin{eqnarray}
        0 &=& {\mbar(E)-m},\nonumber \\
        0 &=& \SI(E,h,m,n),    \nonumber \\
\Df{h}{t} &=& {(\hbar(E)-h)}/{\tau_{h}(E)},\nonumber \\
\Df{n}{t} &=& {(\nbar(E)-n)}/{\tau_{n}(E)},
							\label{hh-slow}
\end{eqnarray}
The finite equations define the slow manifold and the
differential equations define the movement along it.

\Fig{hhtikh}(c) shows a selected trajectory of system \Eq{hh-22}
corresponding to a typical AP solution.  The only
equilibrium of the full system \Eq{hh-22}, corresponding to the
resting state, is at the lower, ``diastolic'' branch of the slow
manifold. If the initial condition is in the basin of the upper
branch, the trajectory starts with a fast initial transient,
corresponding to the upstroke of the AP, then proceeds
along the upper, ``systolic'' branch of the slow manifold, which
corresponds to the plateau of the AP.  When the
trajectory reaches the fold line, a boundary of the systolic branch,
the plateau stage is over. The fast subsystem is no longer bistable.
The only stable equilibrium is now at the diastolic branch.  So the
trajectory jumps down.  This is repolarization from the AP
and it happens at the time scale $t\sim\epsilon$. The
trajectory then slowly proceeds along the diastolic branch towards the
resting state.

So, an inevitable feature of the asymptotics \Eq{hh-22} is 
that the
solution at smaller $\epsilon$
has not only a faster upstroke, but also a faster repolarization, and the
asymptotic limit of the AP shape is rectangular as
opposed to the triangular shape in the exact model, see \fig{hhtikh}(d). This is
undesirable as it means that asymptotic formulae obtained in
this way produce qualitatively inappropriate results.

The practical importance of this excercise is limited as in HH the
AP are not much longer than upstrokes.

\paragraph{Noble 1962.} 
This model is more relevant to cardiac AP. 
The speed analysis of N62 model, similar to the one we have done for HH model,
reveals a different asymptotic structure but ultimately similar results.
\Fig{n62tikh}(a) demonstrates three rather than two different time
scales. Variable $m$ is the fastest of all, we call it
``superfast''. Of the remaining three, variables $E$ and $h$ are fast
and variable $n$ is slow. So we need two small
parameters, 
$\epsilon_1$ to describe the difference
between the fast and superfast time scales,
and
$\epsilon_2$ to describe the difference between the slow
and fast time scales.
System \Eq{hh+n62} then
takes the form
\begin{eqnarray} 
\epsilon_2\epsilon_1 \Df{m}{t} &=& {(\mbar(E)-m)}/{\tau_m(E)},	\nonumber \\
\epsilon_2           \Df{E}{t} &=& -\SI(E,h,m,n)/\CM,		\nonumber \\
\epsilon_2           \Df{h}{t} &=& {(\hbar(E)-h)}/{\tau_h(E)},	\nonumber \\
                     \Df{n}{t} &=& {(\nbar(E)-n)}/{\tau_n(E)}.
								\label{n62-121}
\end{eqnarray}
Consider first the limit $\epsilon_1\rightarrow0$. The superfast
subsystem consists of one differential equation for $m$. It always has
exactly one equilibrium which is always stable. So after a supershort
transient, $m(t)$ is always close to its quasi-stationary value
$\mbar(E(t))$. Thus, with an error $\sim\epsilon_1$ we may approximate
$m$ by $\mbar$ and discard the first equation,
i.e. \emph{adiabatically eliminate} superfast variable $m$.

\myfigure{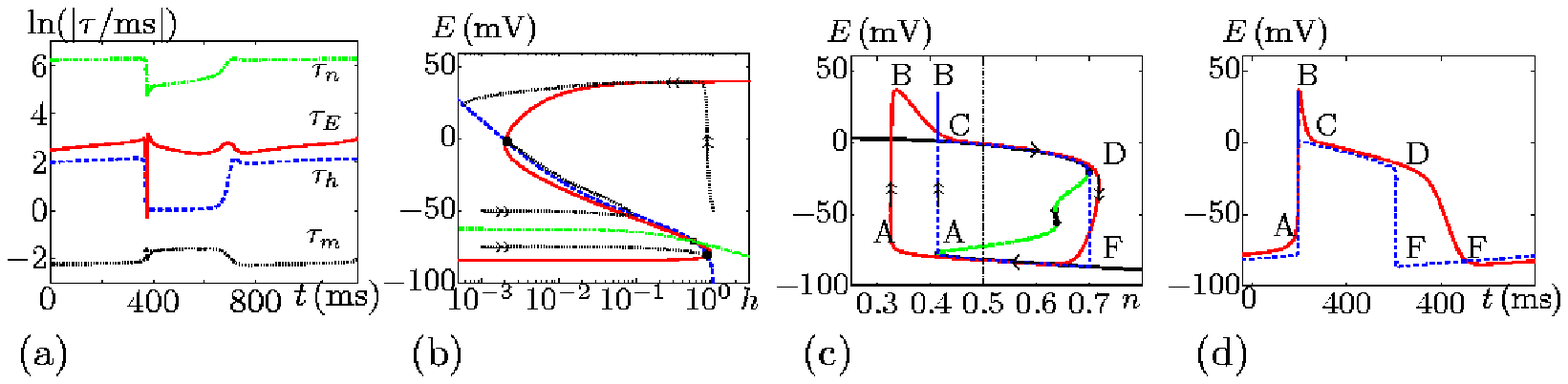}{
Tikhonov asymptotics of N62.
(a,b,d) Notations similar to \fig{hhtikh}.
(c) A two-dimensional view of the stable (black solid) 
and unstable (green dashed) branches of the slow manifold,  and 
typical pacemaker potential trajectories (the limit cycles): 
solid red ($\epsilon_2=1$) and dashed blue ($\epsilon_2=10^{-3}$),
vertical dash-dotted line shows value $n=0.5$ for the fast phase portrait on (b). 
See also \cite{Suckley-Biktashev-2003}.
}{n62tikh}

With the remaining system of three differential equations for $E$, $h$
and $n$, we consider the change of the time variable as before,
$t=\epsilon_2T$, and proceed to the limit $\epsilon_2\rightarrow0$. This
produces the fast subsystem in the form
\begin{eqnarray} 
\Df{E}{T} &=& -\SI(E,h,\mbar(E),n)/\CM,	\nonumber \\
\Df{h}{T} &=& (\hbar(E)-h)/{\tau_h(E)} .
						\label{n62-fast}
\end{eqnarray}
\Fig{n62tikh}(b) shows a phase portrait of this system for a selected
value of $n$ when \Eq{n62-fast} is bistable. As there is one slow variable, all the equilibria of the
fast subsystem form a one-dimensional manifold, 
i.e. a curve, in the three-dimensional phase
space with coordinates $(E,h,n)$. Its projection 
on the plane $(h,E)$ is shown in \fig{n62tikh}(c).  Again
the stable equilibria correspond to one slope (negative for the given
choice of coordinates) and the opposite slope corresponds to the
unstable equilibria. The branches are separated from each other by
fold points. Again we have an upper, systolic branch,
separated from the lower, diastolic branch, and the system has no
alternative but to jump from one branch to the other in the
time scale $t\sim\epsilon_2$. As it happens, there are no true
equilibria in the N62 model at standard parameter values, so
these jumps happen periodically, producing pacemaker potential. This
corresponds to the automaticity of cardiac Purkinje cells. 
Certain physiologically feasible changes of parameters may produce
an asymptotically stable equilibrium at the diastolic branch, i.e turn
an automatic Purkinje cell into an excitable cell. 

The systolic branch is separated from the diastolic
branch, so in the asymptotic limit
$\epsilon_2\rightarrow0$, if the upstrokes are
fast, the repolarizations are
similarly fast. This is in contradiction with the 
behaviour of the full
model, which makes such an asymptotic analysis unsuitable.

\section{Non-Tikhonov asymptotics}
\label{sec:Non-Tikhonov}

\paragraph{Noble 1962.}

To overcome the difficulties of the Tikhonov approach, we
have developed an alternative based on actual
biophysics behind N62 and other cardiac excitation
models, see \fig{n62nont}(a).  The upstroke of an AP is much faster
than the repolarization as the two processes are caused by different
ionic currents.  The upstroke is very fast as the fast Na current causing
it is very large. However all other currents,
including the outward K currents that bring about repolarization
are not as large, and there is no reason to tie the two classes of
currents together in an asymptotic description.  Mathematically, this means that
in the right-hand side of the equation for $E$, only the term
corresponding to the fast Na current is large and should have the
coefficient $\epsilon_2^{-1}$ in front of it.

Next, the fast Na current is large only
during AP upstroke,  and remains small during
other stages. 
The current is
regulated by two gating variables, $m$ and $h$, and the
quasi-stationary value of the specific conductivity of the current,
defined as $W(E)=\mbar^3(E)\hbar(E)$, is always much smaller than one.
This happens because $m,h\in[0,1]$ and $\mbar^3(E)$ is very small for $E$ below a
certain threshold voltage $\Em$, and $\hbar(E)$ is very small for $E$
above another threshold voltage $\Eh$, and $\Eh<\Em$. So the ranges of
almost complete closure of $\mbar^3(E)$ and $\hbar(E)$ overlap.  So
whenever $E$ changes so slow that $m$ and $h$ have enough time to
approach their quasi-stationary values, the fast Na channels are mostly
closed. The possibility for the opening of a large fraction of Na
channels only exists during the fast upstroke, as $m$ gates are
much faster than $h$ gates and have time to open before $h$ close.

Thus, the facts that $m$ gate is much faster than $h$ gate, 
$\mbar^3(E)\ll1$ for $E<\Em$, $\hbar(E)\ll1$ for $E>\Eh$ and $\Eh<\Em$,
are all related and reveal why the 
upstroke of the AP is much faster than 
all other stages.

\myfigure{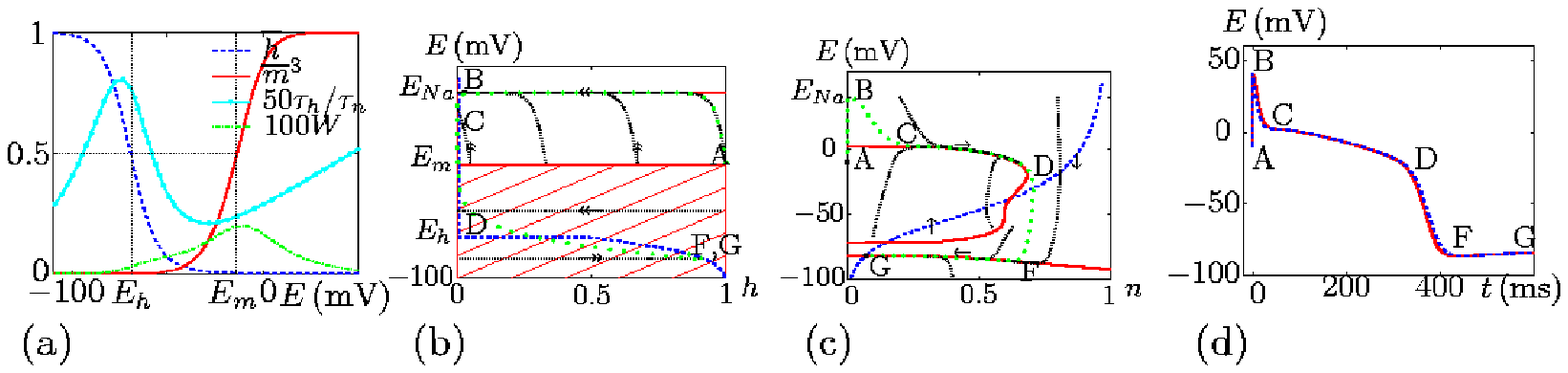}{
Non-Tikhonov asymptotics of N62, excitable variant.
(a) Functions of $E$ illustrating the small quantities taken into account. 
(b) Phase portrait of the fast subsystem. 
(c) Phase portrait of the slow subsystem. 
(d) Action potential solution of \Eq{n62nt-e2} for $\epsilon_2=1$, 
solid red, and for $\epsilon_2=10^{-3}$, dashed blue.
See also \cite{Biktashev-Suckley-2004}.
}{n62nont}

We adopt the
hierarchy of times suggested by the formal speed analysis in the
previous section, i.e. $m$ is a superfast variable, $E$ and $h$ are
equally fast variables and $n$ is a slow variable. We keep the same
notation $\epsilon_1$ and $\epsilon_2$ for the corresponding small parameters.
The small parameters should also ensure that
$\mbar^3(E)$ and $\hbar(E)$ are small in some ranges of $E$ but not in others. 
We identify this smallness with $\epsilon_2$ 
rather than $\epsilon_1$, as it is to compensate the large value
of $\gNa$ outside the upstroke, and the large value of $\gNa$ is described
by $\epsilon_2$. We denote the $\epsilon_2$-dependent versions of
$\mbar^3(E)$ and $\hbar(E)$ as $\emb{\mbar^3}(E;\epsilon_2)$ and $\emb{\hbar}(E;\epsilon_2)$.
In this way we obtain 
\begin{eqnarray}
\epsilon_1\epsilon_2\Df{m}{t} &=& {(\emb{\mbar}(E;\epsilon_2)-m)}/{\tau_{m}(E)}, 
	\qquad \emb{\mbar}(E;0)=M(E)\Heav(E-E_{m}), 	\nonumber\\
\CM\Df{E}{t} &=& 
  \epsilon_2^{-1} \gNa\left(\ENa -E\right) m^3 h 
  +   \gK \left(\EK -E\right) n^4  		\nonumber\\
  &+& \gNai(E)\left(\ENa -E\right) 
  +   \gKi(E)\left(\EK -E\right) 
  +   \gl \left(\El -E\right) ,			\nonumber\\
\epsilon_2\Df{h}{t} &=& {(\emb{\hbar}(E;\epsilon_2)-h)}/{\tau_{h}(E)}, 
	\qquad \hbar(E;0)=H(E)\Heav(E_{h}-E), 	\nonumber\\
\Df{n}{t} &=& {(\nbar(E)-n)}/{\tau_n(E)} ,
						\label{n62nt-e1e2}
\end{eqnarray} 
where $\Heav()$ is the Heaviside function.

The limit $\epsilon_1\rightarrow0$ gives
adiabatic elimination of the $m$ gate, $m\approx\mbar(E)$. 

The analysis of the limit $\epsilon_2\rightarrow0$ is complicated by
a feature incidental to N62 and not found in other
cardiac excitation models. The small conductivity of the window
current, $W(E)$, is multiplied by a large factor $\gNa$.  In N62
the resulting window component of $\INa$ is comparable
to other small currents and cannot be neglected 
outside the upstroke. The implications of this
complication are analysed in \cite{Biktashev-Suckley-2004}. The result, in
brief, is that the $\epsilon_2$-dependent part of \Eq{n62nt-e1e2} can,
in the limit $\epsilon_2\rightarrow0$, 
be replaced with a ``modified N62'' model:
\begin{eqnarray}
\CM\Df{E}{t} &=& 
  \epsilon_2^{-1} \gNa\left(\ENa -E\right) M^3(E)\Heav(E-E_{m}) h 
  +   \gK \left(\EK -E\right) n^4 + G(E)  		\nonumber\\
\epsilon_2\Df{h}{t} &=& {(H(E)\Heav(E_{h}-E)-h)}/{\tau_{h}(E)}, \nonumber\\
\Df{n}{t} &=& {(\nbar(E)-n)}/{\tau_n(E)} ,
						\label{n62nt-e2}
\end{eqnarray} 
where 
$G(E)=\gNa\left(\ENa -E\right) W(E) + \gNai(E)\left(\ENa -E\right) + \gKi(E)\left(\EK -E\right) + \gl\left(\El -E\right)$, and
as before $M(E)\approx\mbar(E)$ for $E>\Em$,
$H(E)\approx\hbar(E)$ for $E<\Eh$ and $W(E)=\mbar^3(E)\hbar(E)$.

The limit
$\epsilon_2\rightarrow0$ of \Eq{n62nt-e2} in the fast time $T=t/\epsilon_2$
gives the fast subsystem
\begin{eqnarray}
\CM\Df{E}{T} &=& \gNa\left(\ENa -E\right) M^3(E)\Heav(E-\Em) h  \nonumber\\
\Df{h}{T} &=& {(H(E)\Heav(\Eh-E)-h)}/{\tau_{h}(E)} .
						\label{n62nt-fast}
\end{eqnarray} 
As intended, \Eq{n62nt-fast} takes into
account only the fast sodium current and the gates controlling
it, and everything else is a small perturbation 
on this timescale.
The phase portrait of \Eq{n62nt-fast} is unusual, see \fig{n62nont}(b).
The horizontal isocline (the red set)
is not just a curve but contains a whole domain $E<E_m$. The
vertical isocline (the blue set) lies entirely within the red
set, 
so the whole line $h=H(E)\Heav(\Eh-E)$ consists
of equilibria. 
An upstroke trajectory may end up in any of the equilibria above
$\Em$, so the height of an upstroke depends on initial conditions.
For subthreshold initial condition, voltage remains unchanged in the
fast time scale. Exactly what happens at the threshold $E=\Em$ depends
on details of approximating function $M(E)$, but in any case it does
not involve any unstable equilibria.  This is all different from
Tikhonov systems (see the paragraph after equation \Eq{hh-fast}) where the height of the
upstroke is fixed, subthreshold potential decays in the fast time
scale and the threshold consists of unstable equilibria and, if appropriate, their
stable manifolds. In this sense, asymptotics of
\Eq{n62nt-fast} give a new meaning to the notion of
excitability, completely different from that in the Tikhonov systems.

Let us consider the slow subsystem of \Eq{n62nt-e2}.
For any value of $n$ we have a whole line of equilibria in
the fast system $h=H(E)\Heav(\Eh-E)$. The
collection of such lines makes a two dimensional manifold in the
three-dimensional space with coordinates $(E,h,n)$. 
So the fast variable $h$ can be adiabatically eliminated on the time scale
$t\sim1$. Thus the slow subsystem, i.e. the limit $\epsilon_2\rightarrow0$ 
in \Eq{n62nt-e2}, is
\begin{eqnarray}
\CM\Df{E}{t} &=& \gK \left(\EK -E\right) n^4 + G(E) ,	\nonumber\\
\Df{n}{t} &=& {(\nbar(E)-n)}/{\tau_n(E)} .
\end{eqnarray}
The phase portrait of this system is shown in
\fig{n62nont}(c). Further discussion of its properties can be found in
\cite{Biktashev-Suckley-2004}. Notice that voltage $E$ features in both the fast and
slow subsystems, i.e. it is a fast or a slow variable depending 
on circumstances. This kind of behaviour is not allowed in Tikhonov
asymptotic theory, so it is ``a non-Tikhonov'' variable.

\paragraph{Courtemanche et al. 1998.}

CRN is a system of 21 ODE modelling electric excitation of human
atrial cells, see \cite{Courtemanche-etal-1998} for a description.

Formal analysis of the time scales $\tau_{\i}$ of dynamic variables
by the same method as we used for HH and N62, reveals a complicated
hierarchy of speeds, which changes during the course of the AP
(see \fig{crnnont}(a)).
From variables with smaller $\tau$'s, we 
select those that remain close to their quasi-stationary
values during an AP, and which can be replaced by those
quasi-stationary values without significantly affecting the AP
solution. We call them supefast variables.  These include $m$, $\ua$
and $w$.  As before, we denote the associated small parameter
$\epsilon_1$.

\myfigure{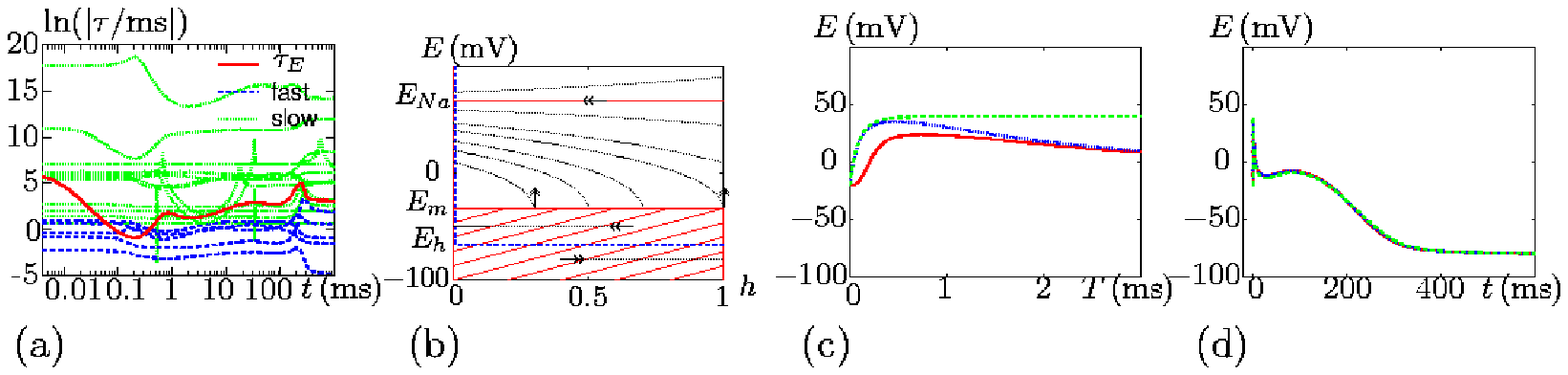}{
Non-Tikhonov asymptotics of the Courtemanche et al. 1998 model.
(a) Characteristic times 
during a typical AP potential solution, as indicated by the
legend. Red solid line: $\tau_E$, blue dashed line:
$\tau$'s of superfast ($m$, $\ua$, $w$) and fast ($h$, $\oa$, $d$),
green dash-dotted lines: $\tau$'s of other, slow variables. 
(b) Phase portrait of the fast subsystem \Eq{CRNfast}.
(c) Action potential upstroke,
$\epsilon_1=\epsilon_2=1$ (solid red),
$\epsilon_1=10^{-3}$, $\epsilon_2=1$ (dotted blue),
$\epsilon_1=\epsilon_2=10^{-3}$ (dashed green),
in the fast time $T=t/\epsilon_2$.
(d) Same, for the whole AP in the slow time $t$.
See also \cite{Suckley-2004}.
}{crnnont}

Next, we identify the fast variables with speeds comparable
to the AP upstroke. This is also done by
comparing the instantaneous values of the variables with the corresponding
quasi-stationary values, and checking how their adiabatic elimination
affects the AP, for the AP solution
\emph{after} the initial upstroke. In this way, we identify variables
$h$, $\oa$ and $d$ as fast, with associated small
parameter $\epsilon_2$.

Similar to N62, the transmembrane voltage is
$\epsilon_2^{-1}$-fast only during the AP upstroke due to
$\epsilon_2^{-1}$-large values of $\INa$ during that period, and is
slow otherwise. This is due to nearly perfect switch behaviour of
the gates $m$ and $h$. The definition of $\INa$ in this model is more
complicated as there is also the $j$ gate; however $j$ is slow
and does not change noticeably during the upstroke.

These considerations lead to the following parameterization of the model:

\begin{eqnarray}
\CM\Df{E}{t} &=& - \left(
  \epsilon_2^{-1}\INa(E,m,h,j) + \SIsmall(E,\dots)
  \right) , 							\nonumber\\
\epsilon_1\epsilon_2\Df{m}{t} &=& {(\emb{\mbar}(E;\epsilon_2)-m)}/{\tau_{m}(E)}, 
	\qquad \emb{\mbar}(E;0)=M(E)\Heav(E-E_{m}), 		\nonumber\\
\epsilon_2\Df{h}{t} &=& {(\emb{\hbar}(E;\epsilon_2)-h)}/{\tau_{h}(E)}, 
	\qquad \emb{\hbar}(E;0)=H(E)\Heav(E_{h}-E), 		\nonumber\\
\epsilon_1\epsilon_2\Df{\ua}{t} &=& {(\uabar(E)-\ua)}/{\tau_{\ua}(E)}, 
								\nonumber\\
\epsilon_1\epsilon_2\Df{w}{t} &=& {(\wbar(E)-w)}/{\tau_{w}(E)}, 
								\nonumber\\
\epsilon_2\Df{\oa}{t} &=& {(\oabar(E)-\oa)}/{\tau_{\oa}(E)}, 
								\nonumber\\
\epsilon_2\Df{d}{t} &=& {(\dbar(E)-d)}/{\tau_{d}(E)}, 
								\nonumber\\
 &\dots&						\label{CRN}
\end{eqnarray}
where $\INa(E,m,h,j)=\gNa m^3hj(E-\ENa)$ is the fast Na current and
$\SIsmall(E,\dots)$ represents all other currents.
Here we have shown only equations that contain $\epsilon_1$ or $\epsilon_2$.
All other equations are the same as in the original model.

As before, we adiabatically eliminate the superfast variables in the
limit $\epsilon_1\rightarrow0$, and turn $\epsilon_2\rightarrow0$ in
the fast time scale $T=t/\epsilon$. This gives the fast subsystem
\begin{eqnarray}
\CM\Df{E}{T} &=& \gNa j (\ENa-E) M^3(E)\Heav(E-\Em) h, \nonumber\\
\Df{h}{T} &=& {(H(E)\Heav(\Eh-E)-h)}/{\tau_{h}(E)},    \nonumber\\
\Df{\oa}{T} &=& {(\oabar(E)-\oa)}/{\tau_{\oa}(E)}, 	\nonumber\\
\Df{d}{T} &=& {(\dbar(E)-d)}/{\tau_{d}(E)}.		\label{CRNfast}
\end{eqnarray}
Notice that in \Eq{CRNfast} the equations for
$\oa$ and $d$ depend on $E$, but equations for $E$ and $h$ do
not depend neither on $\oa$ nor on $\d$.  Thus, the evolution equations for
$E$ and $h$ form a closed subsystem.  This system
is identical to \Eq{n62nt-fast}, up to the values of parameters,
definitions of the functions of $E$ and the presence of the slow
variable $j$ as the factor at the maximal conductance of the Na
current, $\gNa$. The phase portrait of \Eq{CRNfast}, 
\fig{crnnont}(b), is similar to that of N62, \fig{n62nont}(b). So the peculiar
features of the fast subsystem of N62 are not unique and are found
in many cardiac models, including CRN.

With a view of a practical application of approximation 
\Eq{CRNfast}, it is interesting to test its quantitative accuracy. 
This is illustrated in \fig{crnnont}, panel (c) for the shape of the
upstroke in the fast time $T=t/\epsilon$, and panel (d) for the shape
of the AP in the slow time $t$. We see that the
approximation of the AP is very good in both limits
$\epsilon_1\rightarrow0$ and $\epsilon_2\rightarrow0$,
except for the upstroke: e.g. the peak voltage is overestimated by
about 13\,mV. This is mostly due to the limit $\epsilon_1\rightarrow0$,
i.e. replacement of $m$ with $\mbar(E)$. So the 
accuracy could be significantly and easily improved by
retracting the limit $\epsilon_1\rightarrow0$, which
amounts to inclusion of the evolution equation for $m$ instead of
the finite equation $m=\mbar(E)$. Most of the qualitative analysis
remains valid. 
However, here for simplicity we stick to the less accurate but
simpler case $\epsilon_1=0$.
Notice that
in \fig{crnnont}(b--d), we took $M(E)=1$,
$H(E)=1$; the error introduced by that was small compared to other
errors, particularly the error introduced by $m=\mbar(E)$.

\section{Application to spatially distributed systems}
\label{sec:Spatial}

\paragraph{Fronts.}

We now use the fast Na subsystem of the cardiac excitation
\Eq{CRNfast} to consider a propagation of an excitation front through a
cardiac fibre. 
In one spatial dimension, this requres replacement of
ordinary time derivatives with partial derivatives and adding a
diffusion term into the equation for $E$:
\begin{eqnarray}
\df{E}{t} &=&  J(E) \Heav(E-\Em) h + D\ddf{E}{x}, \nonumber\\
\df{h}{t} &=& {(\Heav(\Eh-E)-h)}/{\tau_{h}(E)},
								\label{CRNfastx}
\end{eqnarray}
where $J(E)=\gNa j (\ENa-E) M^3(E)$ and we have put $H(E)=1$.  We
consider solutions in the form of propagating fronts.
For definiteness, let us assume the fronts propagating
leftwards, so $E(x,t)=E(\xi)$, $h(x,t)=h(\xi)$, $\xi=x+ct$,
$h(-\infty)=1$, $h(+\infty)=0$, $E(-\infty)=\Ealp$ and
$E(+\infty)=\Eomg$. In this formulation, we have three constants
characterizing the solutions, the prefront voltage $\Ealp$, the
postfront voltage $\Eomg$ and the speed $c$. It is not obvious
which combinations of the three parameters admit how
many front solutions. So we have considered a ``caricature'' of
\Eq{CRNfastx} by replacing functions $J(E)$ and $\tau_h(E)$ in it with
constants:
\begin{eqnarray}
\df{E}{t} &=&  J \Heav(E-\Em) h + D\ddf{E}{x}, \nonumber\\
\df{h}{t} &=& {(\Heav(\Eh-E)-h)}/{\tau_{h}},
								\label{caric}
\end{eqnarray}

\myfigure{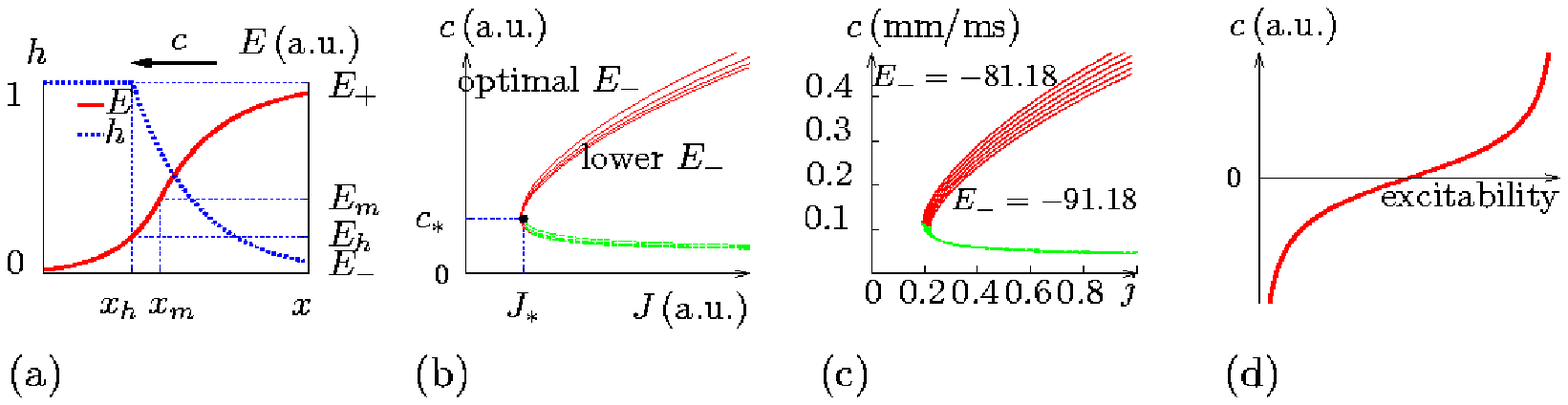}{
Fronts in the spatially distributed Na subsystem.
(a) The structure of the front solution in the caricature model. 
(b) Speed of the fronts as function of excitability, at selected
values of the pre-front voltage, in the caricature model \Eq{caric}.
$J_*$ is minimal excitability at which propagation is possible
at any $E_-$, and $c_*$ is the corresponding propagation speed. 
(c) Speed of the fronts as function of excitability, at selected
values of pre-front voltage, in the Na subsystem of the CRN model
\Eq{CRNfastx}.
On (b) and (c), solid red lines, above and raising, are the stable branches
and dashed green lines, below and decreasing, are the unstable branches. 
(d) For comparison: speed of the fronts in a typical Tikhonov front
(fast susbsystem of the FHN model). 
See also \cite{Biktashev-2002,Biktashev-Biktasheva-2005}.
}{nontwaves}

This system is piecewise linear and admits 
complete analytical investigation. Details can be found in
\cite{Biktashev-2002,Biktashev-2003}; here we only briefly outline the
results. \Fig{nontwaves}(a) illustrates a typical front solution. It
exists if speed $c$ and pre-front voltage $\Ealp$ satisfy a finite
equation involving also the constants $J$ and $\tau_h$.
The resulting dependence of the conduction velocity $c$ on
excitability $J$ for a few selected values of $\Ealp$ is shown in
\fig{nontwaves}(b).  These front solutions exist only for $J$ at or above a
certain minimum $J_{\min}$ which depends on $\Ealp$. For
$J>J_{\min}(\Ealp)$ there are two solutions with different speeds.
Numerical simulations of PDE system \Eq{caric} suggest that solutions
with higher speeds are stable and solutions with lower speeds are
unstable; this has been confirmed analytically by
\citeasnoun{Hinch-2004}.

The replacement of functions $J(E)$ and $\tau_h(E)$ with constants is
a rather crude step.
The purpose of the caricature is not to provide a good
approximation, but to investigate qualitatively the structure of
the solution set. To see if this structure is the same for the more
realistic models, we have solved numerically the boundary-value
problems for the front solutions in \Eq{CRNfastx}. There the role of
the excitability parameter is played by the variable $j$.
The results of the calculations are shown in
\fig{nontwaves}(c).  Not only the topology of the solution set is the
same, but the overall behaviour of $c(j,\Ealp)$ in \Eq{CRNfastx} is
quite similar to that of $c(J,\Ealp)$ in \Eq{caric}, despite 
the crudeness of the caricature. 

PDE simulations show that approximation \Eq{CRNfastx} overestimates
the conduction velocity by almost 50\% compared to the full model,
and the error is again mainly due to the
adiabatic elimination of the $m$ gate.

After eliminating the superfast
variables $m$, $\ua$ and $w$ and the fast variables $h$, $\oa$ and $d$, 
and retaining the non-Tikhonov variable $E$, 
the slow subsystem of \Eq{CRN}
has 16 equations. 
It describes the AP behind the front.

The most important conclusion is that for any
particular value of the prefront voltage $\Ealp$ there is a certain
minimum excitability $\jmin=\jmin(\Ealp)$ and corresponding minimum
propagation speed $\cmin=\cmin(\Ealp)$, and for $j<\jmin$, no steady
front solutions are possible. This is completely different
from the behaviour in FitzHugh-Nagumo (FHN) type systems, 
where local kinetics are Tikhonov and a front is a trigger wave in a
bistable reaction-diffusion system. 
A typical dependence of the speed of such a trigger wave on a slow
variable is shown in \fig{nontwaves}(d): it can be slowed down to a halt
or even reversed. The reversed trigger waves describe backs of
propagating pulses in FHN systems.
Thus, questions about the shape of the backs of APs
and propagating pulses, and the spectrum of propagation speeds of a
propagating excitation wave in a tissue come to be closely related.
In both questions, our new non-Tikhonov approach provides
different answers from the traditional Tikhonov/FHN approach. We have
already seen that the new description is more in line with the
detailed ionic models regarding the back of an AP. In
the next section, we demonstrate the advantage of the new approach
regarding the fronts.

\paragraph{Dissipation of fronts.}

The fast subsystem of a typical spatially-dependent cardiac excitation
model, discussed in the previous section, only provides part of the
answer.  This description should be completed with the description of
the slow movement. 
The fronts
are passing so quickly through every given point that the values of
the slow variables at that point change little while it
happens. 
Away from the fronts, the fast variables keep close to their
quasi-stationary values. In our asymptotics this means, in particular,
that the fast Na channels are closed, and $E$ is not a fast but a slow
variable.  
Assuming absence of spatially sharp inhomogeneities of tissue
properties, simple estimates show that 
outside fronts, the diffusive current is much smaller than ionic currents,
so dynamics of cells there are essentially
the same as dynamics of isolated cells outside  AP upstrokes. 

Propagation of the
next front depends on the transmembrane voltage $E$, which serves
as parameter $\Ealp$, and the slow inactivation
gate $j$ of the fast Na current.
This dependence gives an equation of motion for the front coordinate $x(t)$, 
\begin{equation}
  \Df{x}{t}=c(j(x,t),E(x,t)),
\end{equation}
where the instantaneous speed of the front $c$ is determined by the values
of $E$ and $j$ at the sites through which the front traverses (in case
of $E$ this is the value which would be there if
not the front). This can only continue as long as the function $c(j,E)$
remains defined, i.e while $j(x,t)>\jmin(E(x,t))$. If the
front runs into a region where this is not satisfied, 
its propagation becomes unsustainable. 

\myfigure{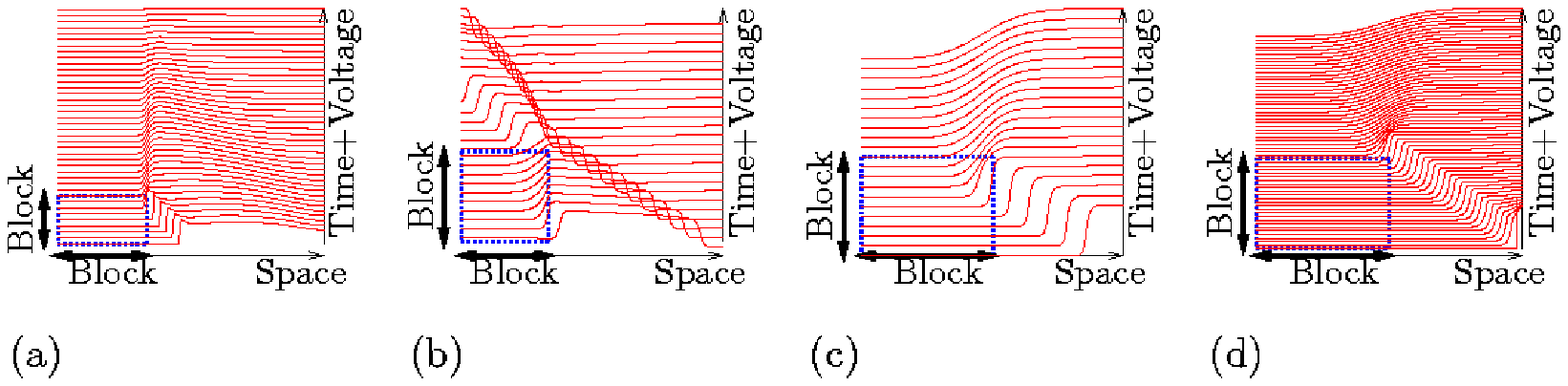}{
Effects of temporary propagation blocks.
(a) In the CRN model.
(b) In the FHN model.
(c) In the caricature Na front.
(d) In the Na subsystem of the CRN model. 
See also \cite{Biktashev-Biktasheva-2005}.
}{nontdiss}

What will happen then is illustrated in \fig{nontdiss}(a), where the
parameter $\gNa$ was varied in space and time.  To make the effect
more prominent, we did not use smooth variation, but put $\gNa=0$ in the
left half of the interval for some time and then restored it to its
normal value. The propagating front reached this region while it was
in the inexcitable state. The result was that the sharp front ceased
to exist, it ``dissipated'', and instead of an active front we observed a
purely diffusive spread of the voltage.  The excitability was restored
a few milliseconds later, but the sharp front did not recover and
diffusive spread of voltage continued, leading eventually to a
complete decay of the wave. Note that the back of the propagating
pulse was still very far when the impact that caused the front
dissipation happened.

This is completely different from the behaviour of a FHN system in
similar circumstances, shown in \fig{nontwaves}(b). 
There propagation was blocked 
for almost the whole duration of the
AP.  And yet when the block was removed, the propagation
of the excitation wave resumed. Only if the block stays so long that
the waveback reaches the block site and the ``wavelength'' reduces to
zero, the wave would not resume. Such considerations have lead to a
widespread, would-be obvious assumption that shrinking of the
excitation wave to ``zero length'' is a necessary condition and
therefore a ``cause'' of the block of propagation of excitation waves
\cite{Weiss-etal-2000}.  Comparison of panels (a) and (b) in
\fig{nontwaves} shows that this is far from true for ionic cardiac
models, where such reduction to zero length happens, but only as a
very distant consequence, rather than a cause, of the propagation
block. The true reason for the block is the disappearance of the fast
Na current at the front, observed phenomenologically as its
dissipation.

We expect that the condition $j>\jmin(E)$ can also serve as a condition of propagation in the
non-stationary situation on the slow space time/scale.
Moreover, we conjecture that
the dissipation of the front will
happen where and when the front runs into a region with
$j<\jmin(E)$. This is illustrated by a simulation shown in
\fig{nontwaves}(c). It is a solution of the caricature system
\Eq{caric}, where the excitability parameter $J$ has been maintained
slightly above the threshold $J_{\min}(\Ealp)$ outside the block
domain, and slightly below it within the block domain.  As a result,
the front propagation has been stopped and never resumed even after
the block has been removed. A similar simulation for the quantitatively
more accurate fast subsystem of the CRN model, \Eq{CRNfastx}, is shown
on panel (d). Both agree with what happens in the full model on panel (a),
and both confirm that the condition $j<\jmin$ is relevant for 
causing front dissipation. 

\paragraph{Break-ups and self-terminations of re-entrant waves}

In two spatial dimensions, the condition $j<\jmin(E)$ may
happen to a piece of a wavefront rather than the whole of it. 
Then instead of a complete block we observe a local block and breakup of the
excitation wave.  This happened in the episode shown in \fig{breakup}.

\myfigure{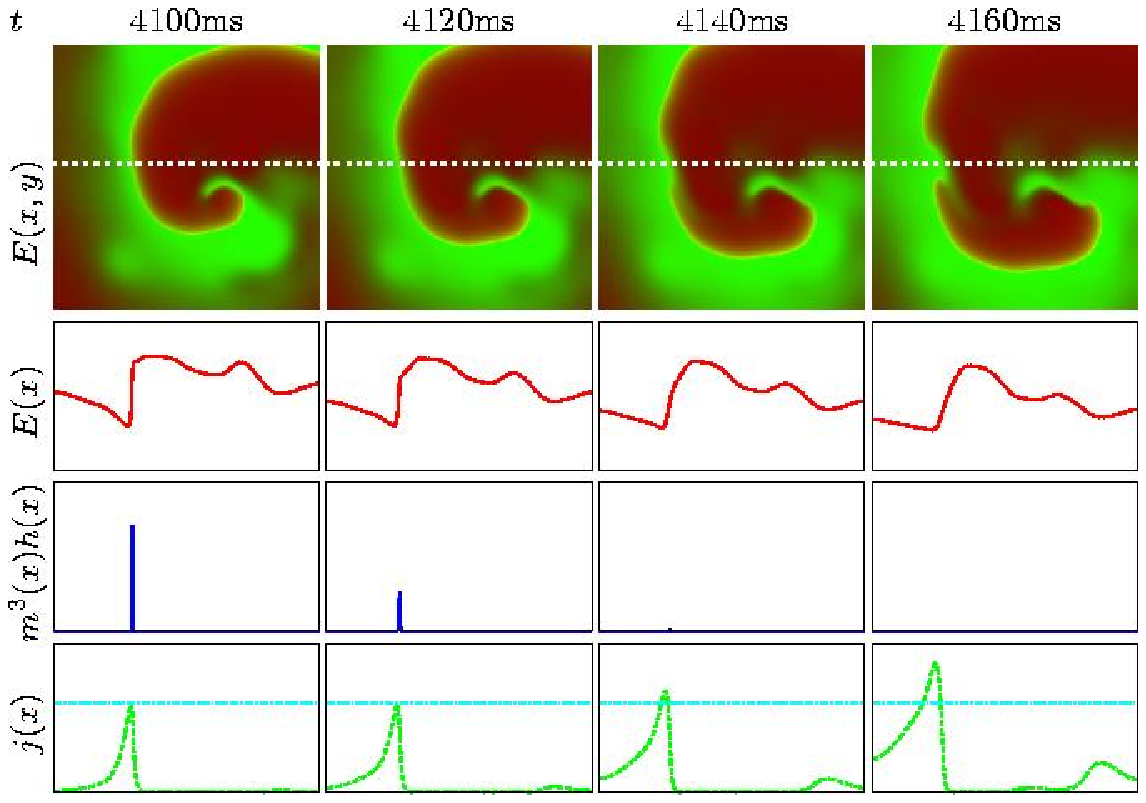}{
  Analysis of a break-up of a re-entrant wave in a simulation similar to
  \fig{selfterm}.
  Top row: snapshots of
  the distribution of the transmembrane voltage, at the selected
  moments of time (designated above the panels). 
  The other three rows:
  profiles of the key dynamic variables (designated on the left) along
  the dotted line shown on the top row panels, at the same moments of
  time. The scale of $E$ is $[-100\,\mV,0\,\mV]$. The scale of $m^3h$ is 
  $[0,0.15]$. The scale of $j$ is $[0,0.5]$.
  Cyan dash-dotted line on $j$ panels represents $\jmin$.
  See also \cite{Biktashev-Biktasheva-2005}.
}{breakup}

The white dotted horizontal line on the top panels goes across the region
where the propagating wave has been blocked and front has dissipated.
The details of how it happened are analysed on the lower three rows,
showing profiles of relevant variables along this dotted white
line. The second row shows the profiles of $E$, which lose the sharp
front gradient after $t=4100\,\ms$. The third row shows the peaks of
the spatial distribution of the product $m^3h$; the sharpness of these
peaks corresponds to the sharp localization of $\INa$ at the front,
and their decay accompanies the process of the front dissipation. The
most instructive is evolution of the profile of the $j$ variable shown
on the bottom row. Consider the column $t=4100\,\ms$.  The gradient of
$j$ ahead of the front, i.e. to the left of the peak of $m^3h$, is
positive, and the front is moving leftwards, i.e. towards smaller
values of $j$.  That is, the front moves into a less excitable area,
left there after the previous rotation of the spiral wave. To the
right of the peak of $\INa$ the gradient of $j$ is negative which
corresponds to the fact that $j$ decreases during the plateau of the
AP. Thus its maximal value at this
particular time is observed at the front. 
This maximal value is, therefore,
the value that should be considered in the condition of the
dissipation, $j<\jmin(E)$. 

As soon as the front has dissipated ($t\approx4120\,\ms$),
the profile of $j$ starts to raise, so the maximum of the $j$ profile
observed at $t=4120\,\ms$ is the lowest one. From the fact that
dissipation has started we conclude that this maximum is below the
critical value $\jmin$.  Assuming that dissipation usually happens soon
after the condition $j<\jmin$ is satisfied (simulations of
\Eq{CRNfastx} show that this happens within a few milliseconds), this
smallest maximum value should be close to $\jmin$, which gives an
empirical method of determining $\jmin$ from numerical simulations of
complete PDE models. For this particular episode the empirical value
of $\jmin$ was found to be approximately 0.3.
This is about 50\% higher than the $\jmin$
predicted for the same range of voltages by \Eq{CRNfastx}; we attribute
this to the approximation $m\approx\mbar$ which caused similar errors
in the upstroke height and front propagation speeds.

\section{Conclusion}

Our new asymptotic approach for cardiac excitability
equations has significant advantages over the traditional approaches. 
The fast subsystem, represented by equations
\Eq{n62nt-fast} and \Eq{CRNfastx}, appears to be typical for cardiac models.
This predicts that front propagation cannot happen at a speed slower
than a certain minimum and at an excitability parameter lower than
a certain minimum. When these conditions are violated the front
dissipates and does not recover even after excitability is
restored. We have obtained a condition for front
dissipation in terms of an inequality involving prefront values of $j$
and $E$. This condition can be used for the analysis of break-up and
self-termination of re-entrant waves in two and three spatial
dimensions.

\paragraph*{Acknowledgments}
This work was supported in part by grants from EPSRC (%
GR/S43498/01, 
GR/S75314/01
) and by an RDF grant
from Liverpool University.


\end{document}